\documentclass[showpacs,amsmath,amssymb,aps,showkeys,floatfix,prd,a4paper]{revtex4}

\usepackage[dvips]{graphicx}
\usepackage{dcolumn}
\usepackage{bm}
\usepackage{epsfig}
\usepackage{amsfonts}
\usepackage{amssymb,amscd}
\usepackage{xcolor}

\newcommand{\rd}{\mbox{\boldmath $\Delta$}}

\def\lsim{\raise0.3ex\hbox{$<$\kern-0.75em\raise-1.1ex\hbox{$\sim$}}}

\def\gsim{\raise0.3ex\hbox{$>$\kern-0.75em\raise-1.1ex\hbox{$\sim$}}}

\newcommand{\be}{\begin{equation}}

\newcommand{\ee}{\end{equation}}

\def\beq{\begin{equation}}

\def\eeq{\end{equation}}

\def\beqa{\begin{eqnarray}}

\def\eeqa{\end{eqnarray}}

\newcommand{\ba}{\begin{eqnarray}}

\newcommand{\ea}{\end{eqnarray}}

\newcommand{\tdm}[1]{\mbox{\boldmath $#1$}}

\newcommand{\rr}{\mbox{\boldmath $r$}}

\newcommand{\bb}{\bm{b}}

\def\gappeq{\mathrel{\rlap {\raise.5ex\hbox{$>$}}

{\lower.5ex\hbox{$\sim$}}}}

\def\lappeq{\mathrel{\rlap{\raise.5ex\hbox{$<$}}

{\lower.5ex\hbox{$\sim$}}}}

\def\Toprel#1\over#2{\mathrel{\mathop{#2}\limits^{#1}}}

\begin{document}

\title{A phenomenological analysis of the nonperturbative QCD contributions \\ for the photon wave function}
\author{V.P. Gon\c{c}alves$^1$ and  B.D.  Moreira$^{1,2}$}
\affiliation{$^1$ High and Medium Energy Group, Instituto de F\'{\i}sica e Matem\'atica,  Universidade Federal de Pelotas (UFPel)\\
Caixa Postal 354,  96010-900, Pelotas, RS, Brazil.\\
$^2$ Departamento de F\'isica, Universidade do Estado de Santa Catarina, 89219-710 Joinville, SC, Brazil.  \\
}

\begin{abstract}
The photon -- induced interactions, present in $ep$, $eA$, $pp$, $pA$, $AA$ and $e^+ e^-$ collisions, are expressed within the color dipole approach in terms of the photon wave function, which describes the transition of the photon into a quark -- antiquark color dipole. Such quantity is usually calculated using perturbation theory assuming that long distance corrections associated to  strong interactions can be neglected.   In this paper we investigate the impact of these nonperturbative QCD (npQCD) corrections to the description of the photon wave function for dipoles of large size in several observables measured at HERA, LEP and LHC. We assume a phenomenological ansatz for the treatment of these npQCD corrections and constrain the free parameters of our model using the experimental data for the photoproduction cross section. The predictions for  the $\gamma \gamma$ cross section, exclusive $\rho$ production in $ep$ collisions and the rapidity distribution for the $\rho$ production in $PbPb$ collisions are compared with the data. We demonstrate that the inclusion of the nonperturbative QCD corrections improves the description of  processes that are dominated by large dipoles.
\end{abstract}

\pacs{12.38.-t, 24.85.+p, 25.30.-c}

\keywords{Quantum Chromodynamics, Photon Wave Function, Photon -- induced interactions.}

\maketitle


\section{Introduction}

\begin{figure}[t]
\begin{tabular}{ccc}
\hspace{-1cm}
{\psfig{figure=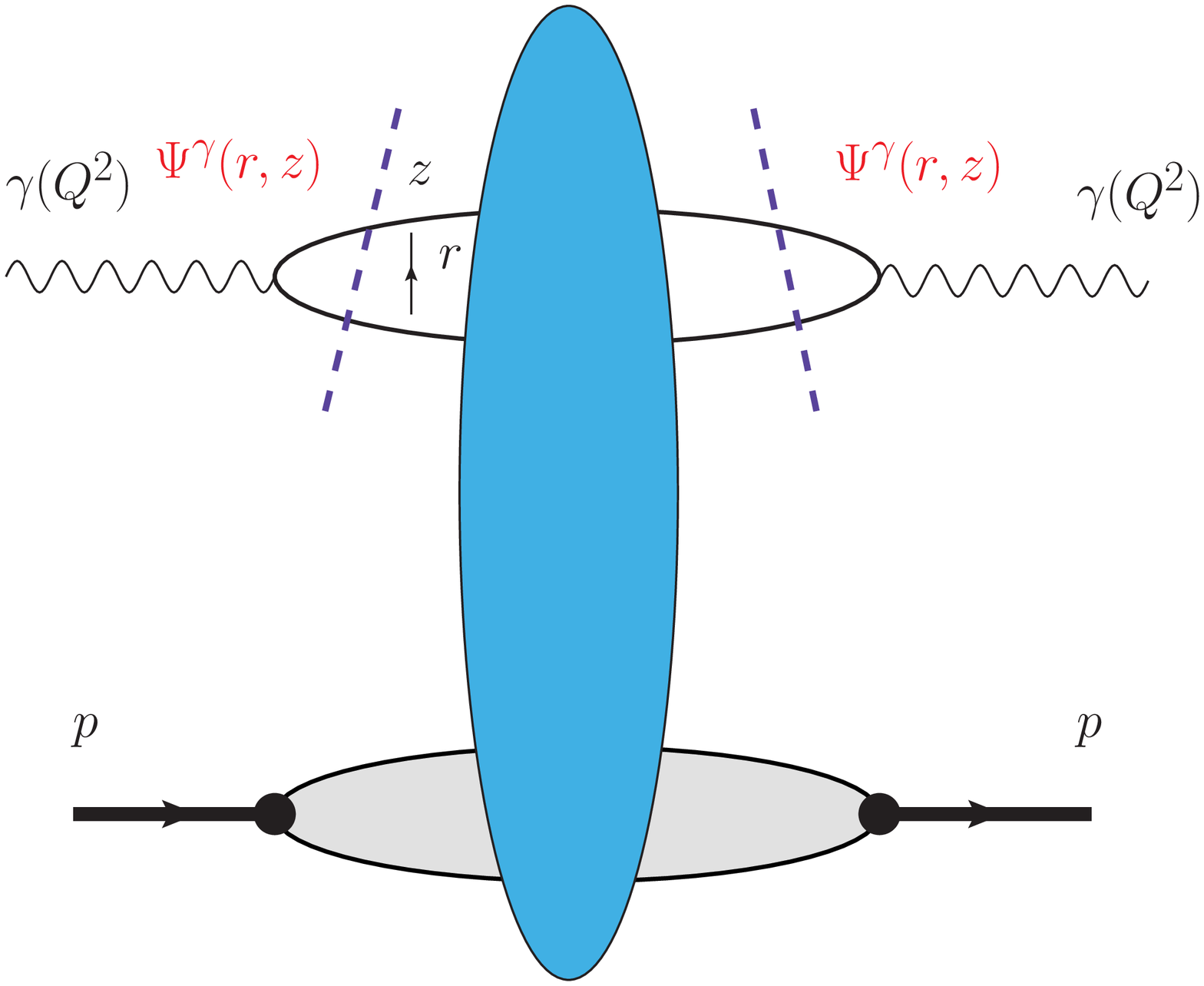,width=5.5cm}}&
{\psfig{figure=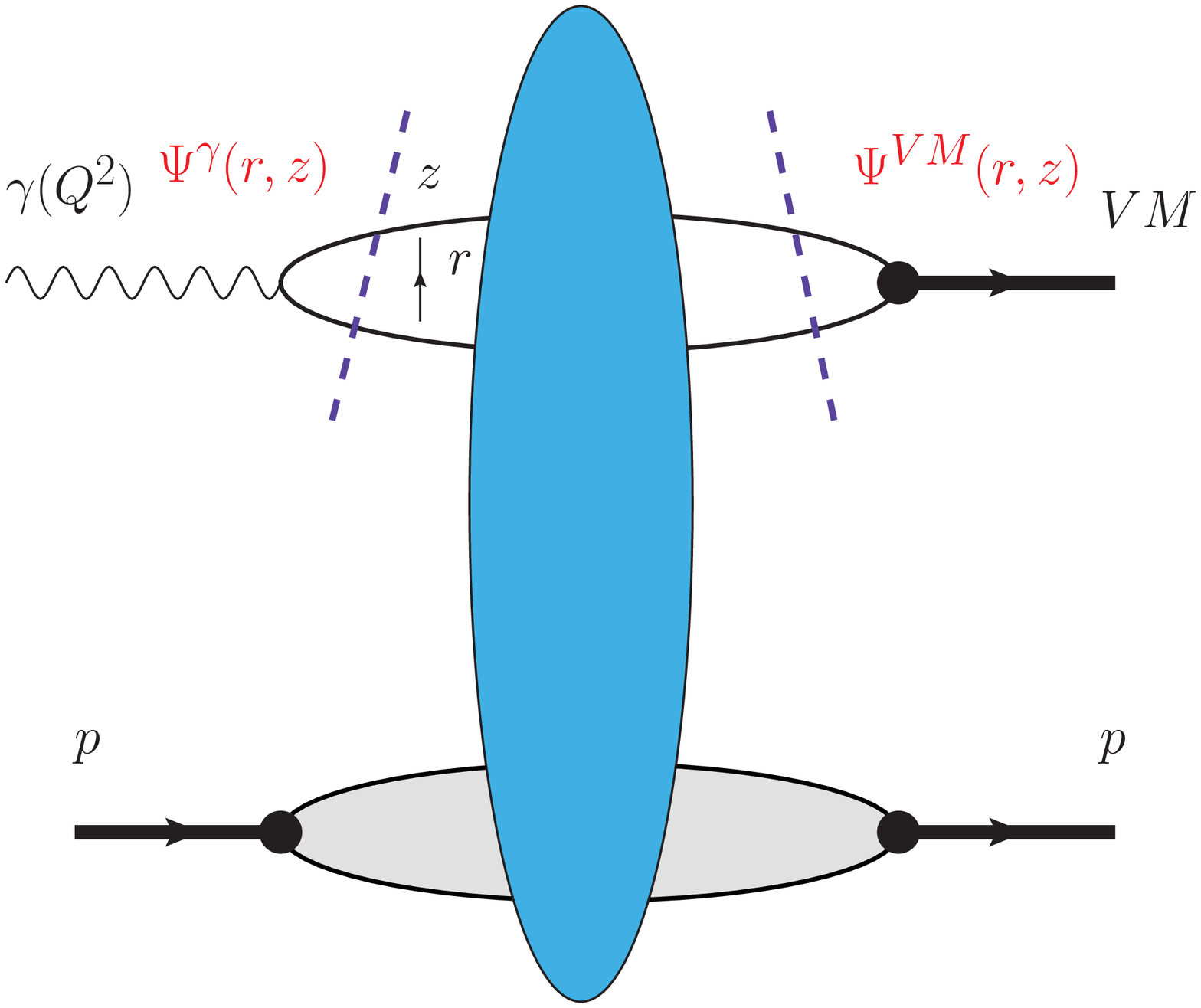,width=5.5cm}} &
{\psfig{figure=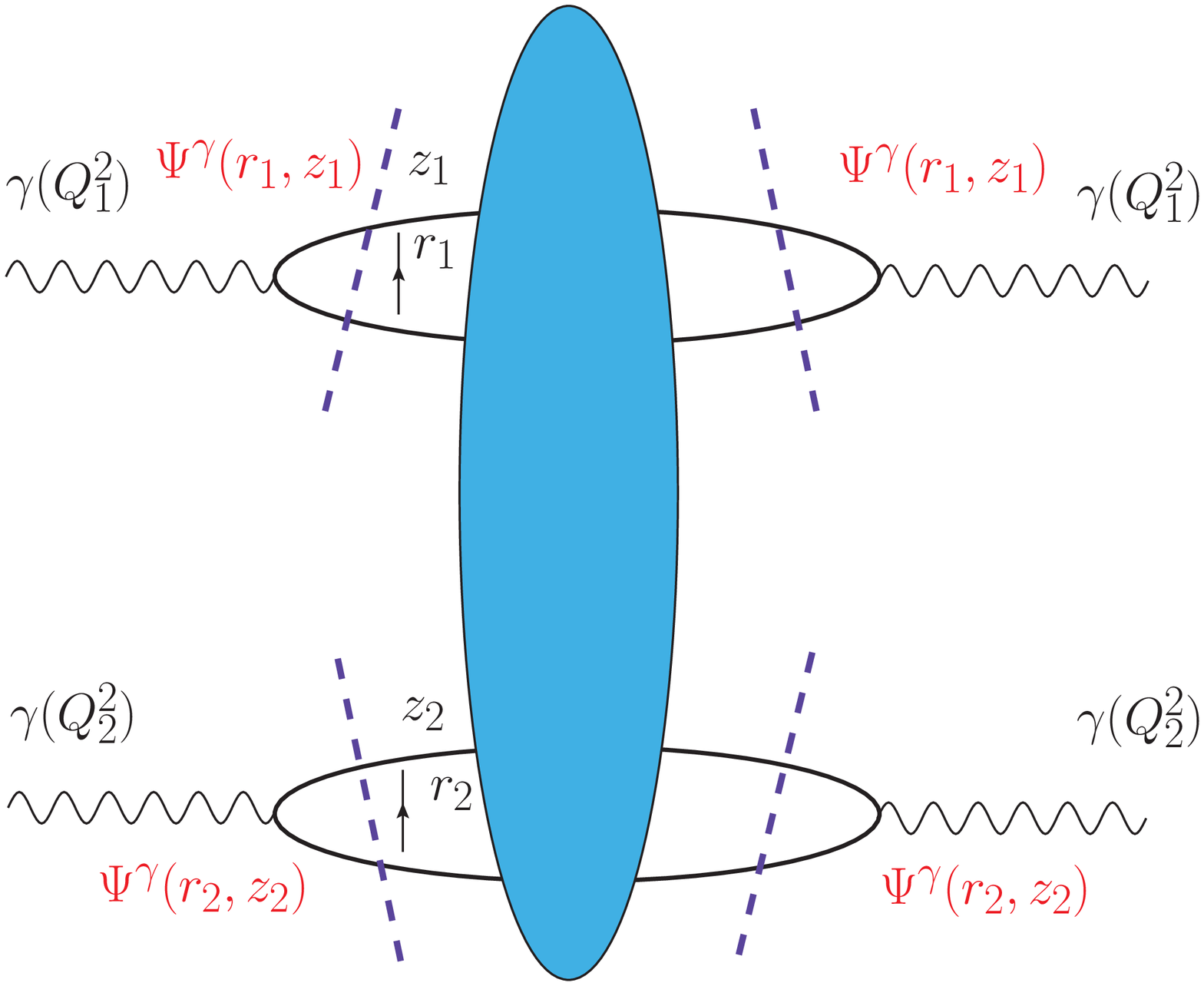,width=5.5cm}} 
 \\
(a) & (b) & (c)  \\ 
\end{tabular}                                                                                                                       
\caption{Diagrammatic representation of the scattering amplitude for the (a) elastic $\gamma^* p$, (b) exclusive $\rho$ meson and (c) $\gamma^* \gamma^*$ processes.}
\label{fig:diagram}
\end{figure}

The description of the high energy behaviour of the strong interactions theory -- the Quantum Chromodynamics (QCD) -- is one of the main challenges of the Standard Model \cite{hdqcd}. Such aspect is directly related to the fact that the physical observables in hadronic collisions  receive contributions of short and long -- distance regimes. For hard processes, as e.g. heavy quark and jet production, the separation between these regimes is usually performed using the collinear factorization theorem \cite{collinear}, which implies that the observables can be estimated in terms of the cross sections for the partonic subprocesses, which can be computed order by order  in perturbation theory, and the parton distribution functions, which are expected to be universal quantities and contain the nonperturbative contributions. 
Such factorization is expected to breakdown at high energies due to the high density of gluons that imply nonlinear corrections to the QCD dynamics \cite{muellerope}. During the last years, distinct approaches have been proposed to generalize the collinear factorization in the high energy regime and take into account of these nonlinear effects \cite{schafer,raju,xiao}. One of these approaches is the color dipole formalism \cite{nik,mueller}, which provides an unified description of inclusive, diffractive  and exclusive observables in electron -- hadron, hadron -- hadron and photon -- photon collisions at high energies. An extensive phenomenology have been performed using this formalism, with the associated predictions being able to successfully describe a large set of experimental data (See e.g. Refs. \cite{gbw,bcgc,ipsat}).
In the particular case of photon -- induced interactions, one of the basic ingredients present in this formalism is the photon wave function, which describes the photon transition into a quark -- antiquark color dipole with a pair separation $\rr$. In Fig. \ref{fig:diagram} we present some examples of photon -- induced processes. For the deep inelastic scattering (DIS) process, one have that the total cross section for the interaction between a virtual photon and the proton can be expressed in terms of the imaginary part of the elastic $\gamma^* p$ amplitude, represented in Fig. \ref{fig:diagram} (a), where $\Psi^{\gamma}(\rr,z)$ is the photon wave function and the blob represents the interaction between the dipole and the proton. Similarly, the scattering amplitude for the exclusive vector meson production  shown in Fig.  \ref{fig:diagram} (b), can also be written in terms of $\Psi^{\gamma}$, with the new ingredient being the vector meson wave function, $\Psi^{VM}(\rr,z)$, which describes the transition of the $q\bar{q}$ dipole into the vector meson. In the case of photon -- photon interactions at high energies, the color dipole formalism predicts that the total cross section can be obtained from the  the elastic $\gamma^* \gamma^*$  amplitude, represented in Fig. \ref{fig:diagram} (c), which depends on the  photon wave functions  and the dipole -- dipole cross section, represented by the blob in the figure.
In general, it is assumed that the photon wave function can be estimated using  perturbation theory, with npQCD corrections being only important to describe the dipole -- proton and dipole -- dipole interactions as well as the vector meson wave function. In other words, it is assumed that the photon wave function does not receive nonperturbative QCD contributions, which can be important for dipoles of large size. Such strong assumption about the photon wave function is fully justified if the observable is dominated by the contribution associated to dipoles of small size, i.e. in processes dominated by short distances between the quark and  antiquark as e.g. the DIS at large $Q^2$, the exclusive production of heavy vector mesons and the interaction between two highly virtual photons.
However,  if the contribution of large size dipoles becomes nonnegligible, it is natural to expect that nonperturbative contributions to the photon wave function will be important and must be taken into account in the description of the observables using the color dipole formalism. 

Our goal in this paper is to perform a phenomenological analysis of the npQCD contributions for the photon wave function, focusing on the description of processes where we expect a larger impact of these corrections as the total photoproduction cross section ($\sigma_{\gamma p}$), the exclusive $\rho$ production  for small values of $Q^2$ in $ep$ collisions and the real $\gamma\gamma$ cross section. In addition, we also will compare our predictions with the data for the exclusive $\rho$ photoproduction in $PbPb$ collisions and for the reduced $ep$ cross section. The experimental data for $\sigma_{\gamma p}$ will be used to constrain the free parameters of our phenomenological model, which implies that the predictions for the other observables will be a direct test of our ansatz as well as of the universality of the color dipole formalism. As demonstrated below, the introduction of a correction factor in the photon wave function, associated to npQCD contributions, improves the description of the experimental data for the observables sensitive to long distances. In addition, we will show that the predictions of the observables dominated by short distances are not modified. The paper is organized as follows. In the next Section, we will present review of the color dipole formalism for the description of photon -- induced interactions. Moreover, our ansatz for the npQCD correction to the photon wave function will be introduced. In Section \ref{sec:res}, {the} free parameters of our model will be constrained using the $\sigma_{\gamma p}$  data and predictions for the $\gamma \gamma$ cross sections and for the exclusive $\rho$ production in $ep$ collisions will be compared with the LEP and HERA data, respectively.
The impact on the exclusive vector meson photoproduction in $PbPb$ collisions will also be discussed. Finally, in Section \ref{sec:conc} we will summarize our main conclusions.

\section{Formalism}

Let us start our
study considering the  description of the deep inelastic $ep$ scattering at high energies. It  is usually described in the
infinite momentum frame  of the hadron in terms of the scattering of
the virtual photon off a sea quark, which is typically emitted  by the
small-$x$ gluons in the proton. However, the photon -- proton scattering can also be described in the dipole frame, in
which most of the energy is carried by the hadron, while the  photon
has just enough energy to dissociate into a quark-antiquark pair
before the scattering \cite{nik}. In this frame, the probing photon
fluctuates into a quark-antiquark pair (a dipole) with transverse
separation $\rr$ long before the interaction, which then scatters off
the proton \cite{nik}. In this approach,  denoted  color dipole formalism,  the total $\gamma^* p$ cross section can be expressed as follows
\begin{eqnarray}
\sigma_{\gamma^* p}(W^2,Q^2)= \sum_{i=L,T} \int dz\, d^2\rr d^2\bb
\,|\Psi^\gamma_{i} (z,\,\rr)|^ 2 \,2 {\cal{N}}(\bb,\rr,x) ,
\label{dipapprox}
\end{eqnarray}
where $W^2$ is the squared photon - hadron center-of-mass energy and $Q^2$ is the photon virtuality,
 The variables  $\rr$ and $z$ are the dipole transverse 
radius and the momentum fraction of the photon carried by a quark (an antiquark carries then $1-z$), respectively, and $\bb$ is the impact parameter of the dipole relative to the proton.
Moreover, 
 $ {\cal N} (x, \rr, \bb)$ is the 
forward dipole-target scattering amplitude  which encodes all the information about the hadronic scattering, 
and thus about the nonlinear and quantum effects in the proton wave function \cite{hdqcd}.
For  the exclusive vector meson production in $ep$ collisions, the color dipole formalism predicts that  the total cross section for the $\gamma^*_{L,T} p \rightarrow E p$ process  can be expressed as follows \cite{ipsat}
\begin{eqnarray}
\sigma_{\gamma^* p \rightarrow E p}(W^2,Q^2) = \sum_{i=L,T} \int dt \,\, 
\frac{d\sigma}{dt}(\gamma^*_{i} p \rightarrow E p)
& = &  \sum_{i=L,T} \int dt \,\, \frac{1}{16\pi}  |{\cal{A}}_{i}^{\gamma^* p \rightarrow E p }(x,Q^2,  \Delta)|^2\,\,, 
\label{dsigdt}
\end{eqnarray}
with $E = \rho, \, J/\Psi,\,...$ and the  amplitude  being given  by
\begin{widetext}
\begin{eqnarray}
 {\cal A}_{L,T}^{\gamma^* p \rightarrow E p }({x},Q^2,\Delta)  =   i
\int \frac{dz}{4\pi} \, d^2\rr \, d^2\bb \,  e^{-i[\bb_h-(1-z)\rr].\rd}  
 \,\, [\Psi^*_{E}(z,\rr) \Psi_{\gamma}(z,\rr,Q^2)]_{L,T}  \,\,2 {\cal{N}}({x},\rr,\bb) 
 \label{amp}
\end{eqnarray}
\end{widetext}
where $[\Psi^*_{E}(z,\rr) \Psi_{\gamma}(z.\rr,Q^2)]$ denotes the wave function overlap between the  virtual photon in the initial state and the vector meson in the final state and $\Delta =  \sqrt{-t}$ is the momentum 
transfer. In this case the cross section also depends on the description of the vector meson wave function $\Psi_{E}$, which describes the transition of the $q \bar{q}$ pair into the vector meson and is sensitive to nonperturbative QCD corrections \cite{vicmagnucmeson}.

As demonstrated e.g. in Ref. \cite{tim}, the color dipole formalism can be directly extended to estimate the total cross section for the $\gamma^{(*)}\gamma^{(*)}$ scattering at high energies, which can be measured in $e^+e^-$ colliders by tagging both outgoing leptons close to the forward direction (For a review see \cite{Nisius99}).
At high energies, the scattering between the two photons can be described in
the dipole frame, in which the photons, with virtualities
$Q_{1,2}^2$, fluctuate into quark-antiquark pairs
(two dipoles) with transverse sizes $\rr_{1,2}$, which then interact and produce
the final state $X$. Within such formalism, the part of
the two-photon total cross section that determines the energy behaviour  at high energies corresponds to the exchange of gluonic degrees of freedom and is given by \cite{DDR}
\begin{equation}\label{eq:gluon}
\sigma^{\gamma \gamma}_{ij}
(W^2,Q_1^2,Q_2^2)=\sum_{a,b=1}^{N_f}\int {\rm d}z_1\int{\rm d}^2
{\bm r}_1|\Psi^{\gamma}_{i,a}(z_1,{\bm r}_1)|^2\int {\rm d}z_2\int{\rm d}^2
{\bm r}_2|\Psi^{\gamma}_{j,b}(z_2,{\bm r}_2)|^2\sigma_{a,b}^{dd}(r_1,r_2,Y).
\end{equation}
In the above formula, $W^2$ is the collision center of mass squared energy, $z_{1,2}$ are the longitudinal momentum fractions of the quarks in the photons, $\Psi^{\gamma}_{i,a}(z_k,\bm{r})$ denotes the photon wave function, the indices $i,\,j$ label the polarisation states of the virtual photons ($i,\,j=$L or T) and $a,\,b$ label the quark flavours. Moreover,  the interaction is described by $\sigma_{a,b}^{dd}(r_1,r_2,Y)$, which is the dipole-dipole
cross section. In the eikonal approximation, it can be expressed by \cite{erikegam}
\be\label{eq:ddeikon}
\sigma^{dd}(\rr_1,\rr_2,Y)=2\int{\rm d}^2\bb\,{\cal{N}}(\rr_1,\rr_2,\bb,Y)
\ee
where ${\cal{N}}(\rr_1,\rr_2,\bb,Y)$ is the imaginary part of the scattering amplitude for two dipoles with transverse sizes $\rr_1$ and $\rr_2$, relative impact parameter $\bb$ and rapidity separation $Y$.
Following Refs. \cite{tim,erikegam}, we will assume that 
\begin{equation}
\sigma^{dd}_{a,b}(r_1,r_2,Y) = \sigma_0^{a,b}\, {\cal{N}}(\rr_{\rm\small  eff},Y=\ln(1/\bar x_{ab}))
\label{sigmadd_mot}
\end{equation}
where 
\begin{equation}
r^2_{\rm\small  eff}\; = \;{r_1^2r_2^2\over r_1^2+r_2^2}\,\,\, \mbox{and}\,\,\, \bar x_{ab} \; = \;{Q_1^2 + Q_2^2 +4m_a^2+4m_b^2\over W^2+Q_1^2+Q_2^2}\,\,.
\label{reff}
\end{equation}
and  $\sigma_0^{a,b} = (2/3) \sigma_0$, with $\sigma_0$ being a free parameter determined  by fitting the DIS HERA data. This relation can be justified by the quark counting rule, as the ratio between  the number of constituent quarks in a photon  and the
corresponding number of constituent quarks in the proton.  In Refs. \cite{tim,erikegam} the authors have demonstrated that the LEP data for the $\gamma \gamma$ interactions can be described by the color dipole formalism if the values of the light quark masses are increased in comparison to those need to describe the HERA data.

The above results demonstrate that in color dipole formalism, the different observables can be fully determined by the modelling of the dipole scattering amplitude and the photon and vector meson wave functions.
At high energies the evolution with the Bjorken -- $x$ variable of
${\cal{N}}(\rr,\bb,x)$  is given by the infinite hierarchy of equations, the so called
Balitsky-JIMWLK equations \cite{bal,cgc}, which reduces in the mean field approximation to the Balitsky-Kovchegov (BK) equation \cite{bal,kovchegov}. It is useful to assume  the translational invariance approximation, which regards hadron homogeneity in the
transverse plane, which implies that  ${\cal{N}}(\rr,\bb,x) = S(\bb) {\cal{N}}(\rr,x)$, 
where $S(\bb)$ is the profile function of the proton, which  sets the normalization.
During the last years, the description of ${\cal{N}}(\rr,x)$ have been a subject of intense activity (See e.g. Refs. \cite{gbw,iim,bcgc,ipsat, runningBK}). As our focus is not the discussion of the QCD dynamics at high energies, but instead to estimate the impact of npQCD corrections in the photon wave function, we will assume in what follows the phenomenological saturation model proposed  in Ref. \cite{gbw}, denoted GBW model hereafter. In this model, the scattering amplitude, ${\cal{N}}(\rr,x)$, is given by \cite{gbw}
\be\label{eq:gbw}
{\cal{N}}(\rr,x)=1-e^{-\rr^2Q_s^2(x)/4},
\ee
where the saturation scale is given by $Q_s^2(x)=Q_0^2\left(x_0/x\right)^{\lambda}$. The parameters $x_0$, $\lambda$ and $\sigma_0 = \int d^2 \bb S(\bb)$ were obtained by fits to the DIS HERA data for $x \le 0.01$ assuming that the light quark masses are $m_u = m_d  = m_s = 0.14$ GeV and that the squared photon wave functions  $|\Psi_T(z,\rr)|^2$ and $|\Psi_L(z,\rr)|^2$ are given by
\begin{eqnarray}
|\Psi^{\gamma}_{T}(z,{\tdm r})|^2\; &  = & \;
{6\alpha_{em}\over 4 \pi^2} \sum_f e_f^2\{
[z^2+(1-z)^2]\;\epsilon_f ^2 K_1^2 (\epsilon_{f}r)
+m_f^2\,K_0^2(\epsilon_{f}r)\}\,\,, 
\\
 |\Psi^{\gamma}_{L}(z,{\tdm r})|^2\; & = & \;
{6\alpha_{em}\over 4 \pi^2} \sum_f e_f^2
 \,[4 Q^2 z^2 (1-z)^2 \; K_0^2 (\epsilon_{f}r)]\,\,,\label{psit}
\end{eqnarray}
with  $(\epsilon_f)^2=z(1-z) Q^2 +
m_f^2$, 
$e_f$ and $m_f$ denote the charge and mass of the quark of flavor $f$ and the 
functions $K_0$ and $K_1$ are the McDonald--Bessel functions.  These squared wave functions are also  used in the phenomenological studies of the photon -- photon interactions using the color dipole formalism \cite{tim,erikegam}. It is important to emphasize that such expressions are derived by calculating $\Psi^\gamma_{i} (z,\,\rr)$ using light -- cone perturbation theory and assuming that the quark and antiquark can be represented by asymptotic states, which are not affected by nonperturbative effects (See e.g. Refs. \cite{nik,ipsat}).

An important aspect in the color dipole formalism is that the observables are obtained by integrating the photon wave function over all possible values of the pair separation $\rr$. As demonstrated in several papers (See e.g. Refs. \cite{gbw,simone,ipsat,nosuphic}), the contribution of large -- $\rr$ is strongly suppressed by nonlinear effects and/or by the presence of a hard scale in the process, which can be the photon virtuality, the quark mass or the mass the vector meson in final state.  
One typical example of a process dominated by small dipoles is the exclusive $\Upsilon$ production at large $Q^2$. On the other hand, the contribution of small and large size dipoles for the exclusive $\rho$ production is strongly dependent on the photon virtuality. The  HERA data for this observable have  demonstrated that for large $Q^2$ the energy dependence of the total cross section is similar to that observed for the $J/\Psi$ production, while for $Q^2 \rightarrow 0$ the energy dependence becomes similar to that expected for a soft process \cite{h1data1,h1data2}. Such behavior is directly associated to the fact that for small values of $Q^2$ the contribution of large size dipoles for the exclusive $\rho$ production becomes nonnegligible. Therefore, we expect that the description of this observable to be sensitive to nonperturbative QCD corrections. Similarly, we also expect a nonnegligible npQCD contribution for the photoproduction cross section, $\sigma_{\gamma^* p}(W^2,Q^2=0)$, and for the real photon -- photon cross section, $\sigma^{\gamma \gamma} (W^2,Q_1^2 = 0,Q_2^2 = 0)$. Such expectation have been previously discussed in some few studies \cite{forshaw,ewerz,lund,berger,kopedvcs}. Inspired by Ref. \cite{berger}, we will assume that the main impact of npQCD contributions for the photon wave function can be modelled by the replacement $\Psi_i^{\gamma} (z,\rr) \rightarrow \sqrt{f_c (\rr)} \,\,\Psi_i^{\gamma} (z,\rr)$, where the correction factor $f_c$ is assumed to be
\begin{eqnarray}
 f_{c}(r) = \left[
 \frac{1 + B \exp\left( -\omega^{2} (r - {\cal R})^{2} \right)}
 {1 + B \exp\left( -\omega^{2} {\cal R}^{2} \right)}
 \right].
\end{eqnarray}
with $B$ and $\omega$ being free parameters and  ${\cal R} = 6.8$ GeV$^{-1}$. Such factor suppress the contribution of large dipoles, as expected from confinement effects which preclude a unrestricted increasing of the distance between quark and antiquark.  In the next Section, we will constrain $B$ and $\omega$ by fitting the experimental data for $\sigma_{\gamma^* p}(W^2,Q^2=0)$ and derive parameter free predictions for $\sigma^{\gamma \gamma} (W^2,Q_1^2 = 0,Q_2^2 = 0)$ and for the exclusive $\rho$ production at HERA and LHC.

\section{Results}
\label{sec:res}
\begin{figure}
\centerline{\psfig{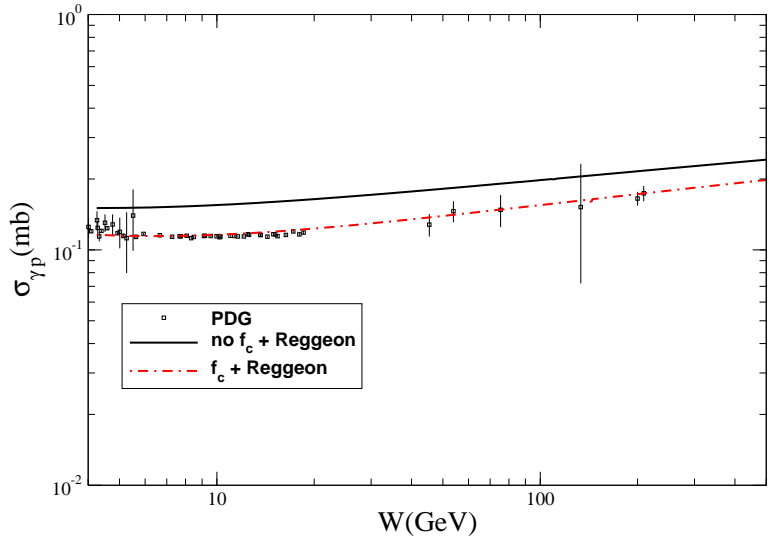}}

\caption{Energy dependence of the photoproduction cross section.}
\label{fig:photo}
\end{figure}

The photoproduction cross section, $\sigma_{\gamma p}(W^2)$, has been measured by several experiments \cite{pdg} at small and large center -- of -- mass energies. At low energies, the energy dependence is expected to be described by a reggeon contribution, which represents a nonperturbative phenomenon related to Reggeon trajectories of light mesons  and is characterized by a decreasing energy dependence. On the other hand, at high energies, the Pomeron contribution, associated to a gluonic exchange, is expected to dominate. Following Ref. \cite{tim}, we will assume that the Pomeron term is described by the color dipole formalism. Consequently, we will have that 
\begin{eqnarray}
\sigma_{\gamma p} (W^2) =  \sigma_{\gamma^* p} (W^2,Q^2=0) \,\,\,[\mbox{Eq. (\ref{dipapprox})}]  +  \sigma_{\gamma p}^R (W^2) \label{gamap} 
\end{eqnarray}
where $\sigma_{\gamma p}^R = A_{\gamma p} .(W^2)^{-\eta}$, with $A_{\gamma p} = 0.12$ mb and $\eta = 0.3$ \cite{tim}. Initially, let's estimate Eq. (\ref{dipapprox}) considering the set of parameters originally derived in Ref. \cite{gbw} using the DIS data  and without the inclusion of the correction factor $f_c$. The resulting prediction for $\sigma_{\gamma p} (W^2)$ is represented by the solid line in  Fig. \ref{fig:photo}, which is  denoted {\it no-fc + Reggeon}. As already emphasized in Ref. \cite{tim}, the original GBW model overestimate the experimental data. In Ref. \cite{tim} the authors proposed to increase the value of the light quark masses in comparison to that need to describe the DIS data. In our approach we will not modify the light quark masses, but instead we will include the factor $f_c$ discussed in the previous Section. In Fig. \ref{fig:photo} we demonstrate that the experimental data can be described including $f_c$ and assuming that $B = -0.9$ and $\omega = 0.15$ GeV.
 
\begin{figure}
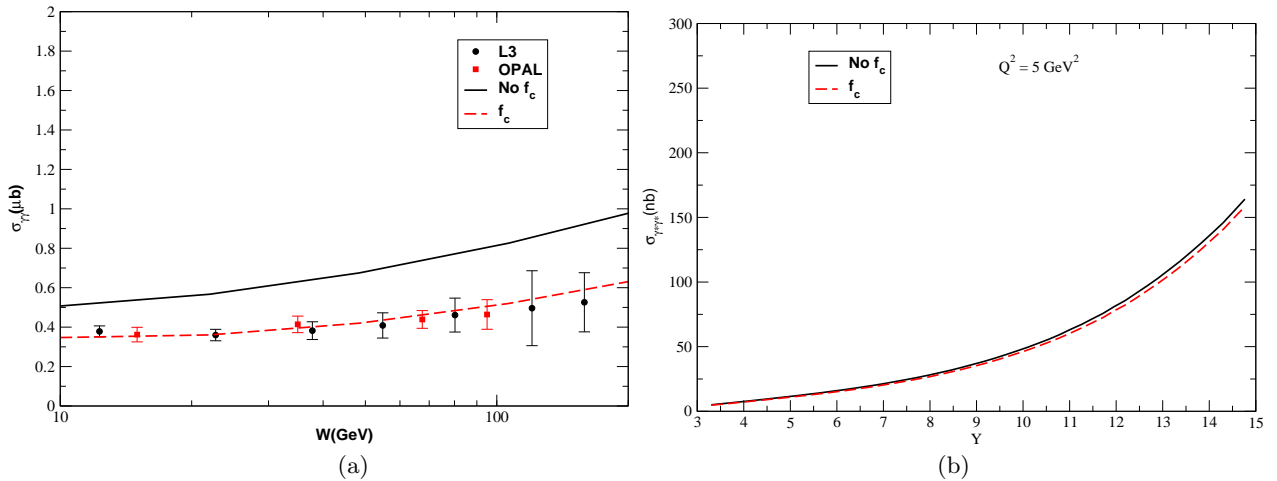

\begin{tabular}{cc}
\hspace{-1cm}
{\psfig{figure=gamma-gamma-REAL.eps,width=8.2cm}}&
{\psfig{figure=virtual_Q2_5_tkm_gbw.eps,width=8.2cm}} 
 \\
(a) & (b)  \\ 
\end{tabular}                                                                                                                       
\caption{(a) Energy dependence of the real $\gamma \gamma$ cross section.  (b) Rapidity dependence of the  virtual $\gamma^* \gamma^*$  cross section for $Q_1^2 = Q_2^2 = 5$ GeV$^2$. Data from Ref. \cite{l3_real}.}
\label{fig:gamagama}
\end{figure}

Let's now estimate the impact of $f_c$ in other observables. In our analysis we will assume the same assumptions considered in  previous studies about photon -- photon interactions and exclusive $\rho$ production, only including the phenomenological factor $f_c$ and assuming that the light quark masses are those used in the description of the DIS data.  Consequently, we will derive parameter free predictions. In Fig. \ref{fig:gamagama} (a) we present our predictions for the energy dependence of the real $\gamma \gamma$ cross section. We have that the inclusion $f_c$ implies that the color dipole formalism is able to describe the experimental data for the collision between two real photons. We also can estimate the  impact of $f_c$ for the collision between virtual photons. As discussed before, for this case the cross section is expected to be dominated by small dipoles and, consequently, to be insensitive to nonperturbative QCD corrections in the photon wave function. Such expectation is confirmed by the results presented in Fig. \ref{fig:gamagama} (b), where we present our predictions for the rapidity dependence, $Y\equiv \ln (W^2/Q_1 Q_2)$, of the virtual $\gamma^* \gamma^*$ cross section for $Q_1^2 = Q_2^2 = 5$ GeV$^2$.

\begin{figure}
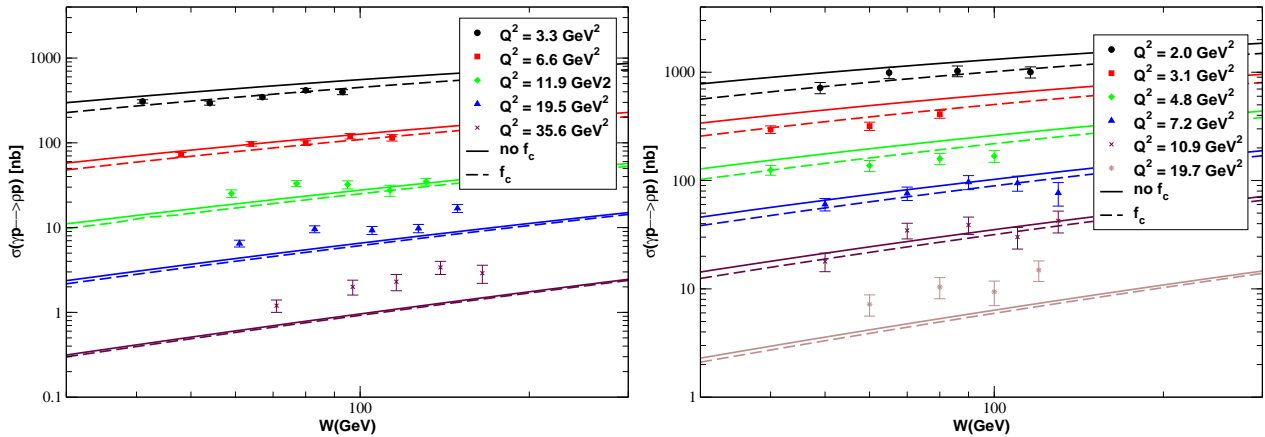

\begin{tabular}{cc}
\hspace{-1cm}
{\psfig{figure=2010_gamma_p_rho_p.eps,width=8.2cm}}&
{\psfig{figure=2000_gamma_p_rho_p.eps,width=8.2cm}} 
\end{tabular}                                                                                                                       
\caption{Energy dependence of the exclusive $\rho$ production cross section for different values of the photon virtuality $Q^2$. Data from Refs. \cite{h1data1,h1data2}.}
\label{fig:rhoep}
\end{figure}

\begin{figure}
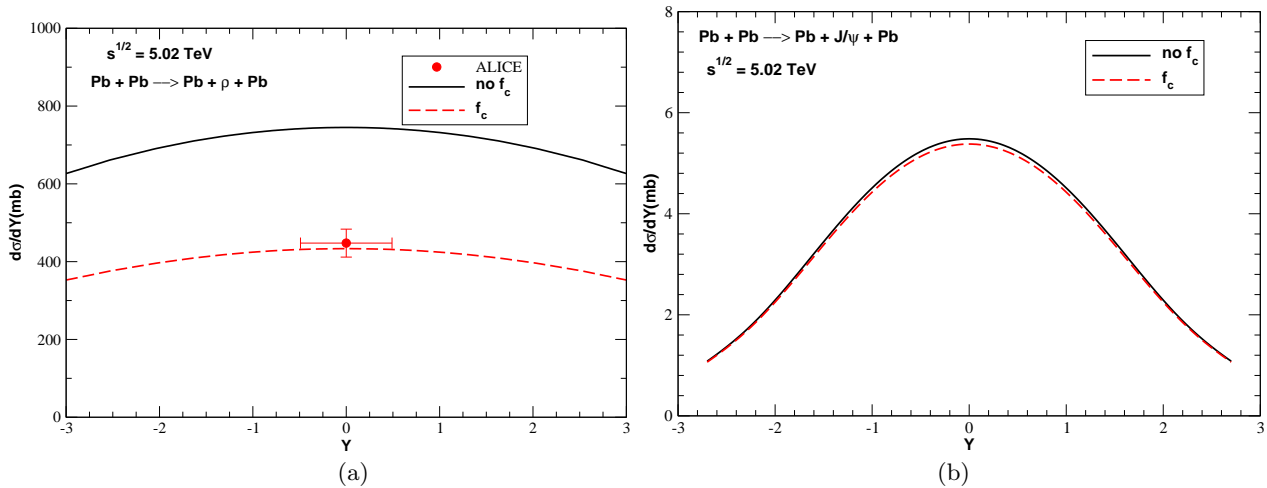

\begin{tabular}{cc}
\hspace{-1cm}
{\psfig{figure=dist_rap_rho2.eps,width=8.2cm}}&
{\psfig{figure=compara_AA_jpsi.eps,width=8.2cm}} 
 \\
(a) & (b)  \\ 
\end{tabular}                                                                                                                       
\caption{Rapidity distribution for the exclusive (a) $\rho$ and (b) $J/\Psi$ photoproduction in $PbPb$ collisions at $\sqrt{s} = 5.02$ TeV. Data from Ref. \cite{alice}.}
\label{fig:rhouphic}
\end{figure}

In Fig. \ref{fig:rhoep} we present our predictions for the exclusive $\rho$ production in $ep$ collisions for different values of the photon virtuality $Q^2$. As in previous studies \cite{nosuphic}, we will consider the Boosted Gaussian model for the  overlap function $[\Psi^*_{\rho}(z,\rr) \Psi_{\gamma}(z.\rr,Q^2)]$. As expected from our previous discussions, we have that the impact of  $f_c$ depends on 
 $Q^2$. For large photon virtualities, it is negligible but becomes important with the decreasing of $Q^2$. Our results indicate that the description of the data for small $Q^2$ is improved by the inclusion of $f_c$. On the other hand, at large $Q^2$ the data is not well described, which is associated to the  fact that the GBW model does not taken into account of the DGLAP evolution. The larger impact of the nonperturbative QCD corrections is expected for the exclusive $\rho$ photoproduction. Such process have been studied in ultraperipheral heavy -- ion collisions at the LHC \cite{nosuphic,vicmag,frank}, which are characterized by an impact parameter larger than the sum of the nuclear radius \cite{uphic}. In Fig. \ref{fig:rhouphic} (a) we present our predictions for the rapidity distribution of the vector meson for $PbPb$ collisions at $\sqrt{s} = 5.02$ TeV, derived following Refs. \cite{nosuphic,bruno1} with and without the inclusion of the correction factor $f_c$. Our results demonstrate that the ALICE data can be described by the GBW model if $f_c$ is included in the calculation. On the other hand, the results presented in Fig.  
 \ref{fig:rhouphic} (b) show that the impact of this correction is small in the exclusive $J/\Psi$ photoproduction.

\begin{figure}
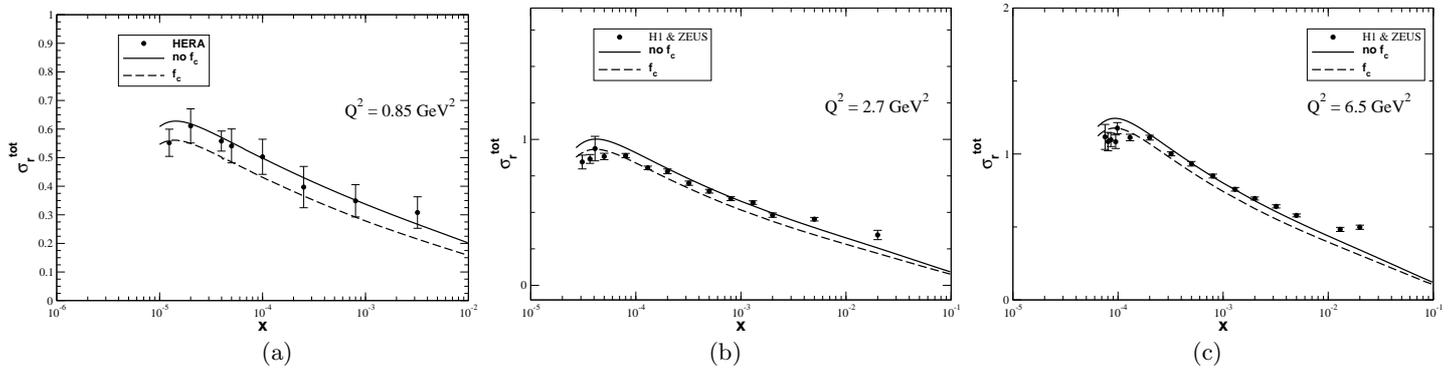

\begin{tabular}{ccc}
\hspace{-1cm}
{\psfig{figure=sigredtot_hera_q2_0_85.eps,width=6.2cm}}&
{\psfig{figure=sigredtot_hera_q2_2_7.eps,width=6.2cm}} &
{\psfig{figure=sigredtot_hera_q2_6_5.eps,width=6.2cm}}
 \\
(a) & (b) & (c) \\ 
\end{tabular}                                                                                                                       
\caption{Bjorken -- $x$ dependence of the reduced $ep$ cross section for different values of the photon virtuality $Q^2$. Data from Ref. \cite{datared}.}
\label{fig:reduced}
\end{figure}

Finally, let's estimate the impact of $f_c$ on the description of the reduced $ep$ cross section for different values of $Q^2$. During the last years, the ZEUS and H1 Collaborations have released very precised data for this quantity, which can be expressed in terms of $\sigma_{\gamma^* p}$ \cite{datared}. In Fig. \ref{fig:reduced} we compare our predictions with the experimental data. As expected, the impact of $f_c$ decreases at large photon virtualities and becomes nonnegligible at small $Q^2$. 
 As the parameters of the GBW model, as well from other phenomenological saturation models, are in general  constrained using the  HERA data  without the inclusion of the $f_c$ correction, these results indicate that one important next step is to perform a new global fit of the dipole -- proton scattering amplitude taking into account of the nonperturbative QCD corrections to the photon wave function. Work in this direction is in progress.

A comment is in order, before to summarize our main conclusions in the next Section. Our results indicated that the inclusion of the nonperturbative factor improves the description of the $\gamma p $ and $\gamma \gamma$ observables that are dominated by long distances considering the same quark mass used in the analysis of the DIS data. One important question is: can we obtain the same improvement adjusting the quark mass and without the inclusion of $f_c$?
We have verified that it is not possible to describe all set of data only modifying the quark mass. We have that the description of the real $\gamma \gamma$ and $\gamma p$ cross section demands a large value of the quark mass ($\approx 0.23$ GeV),  but such value spoils the description of the data for the exclusive $\rho$ production and DIS at medium -- $Q^2$. On the other hand, assuming that $m_f \le 0.14$ GeV, which is preferred by the DIS and $\rho$ data, we were not able to describe the real $\gamma \gamma$ and $\gamma p$ data. Such results point out that  nonperturbative effects must be taken into account in the photon wave function in order to describe, in an unified way, the photon -- induced processes.

\section{Summary}
\label{sec:conc}
In recent years the color dipole formalism has been largely used to estimate inclusive and exclusive observables taking into account of nonlinear effects in the QCD dynamics at high energies. One of the basic ingredients to describe the photon -- induced interactions present in $ep$, $eA$, $\gamma \gamma$ and ultraperipheral heavy ion collisions is the photon wave function. In general, this quantity is estimated using perturbation theory and neglecting long distance corrections associated to nonperturbative QCD effects. In this paper, we have performed a phenomenological analysis of these corrections in a set of observables that are sensitive to dipoles of large size. Assuming the GBW model to describe the dipole -- proton scattering amplittude, we have fixed the two free parameters of our model using the experimental data for the photoproduction cross section and obtained parameter free predictions for the  photon -- photon  and exclusive $\rho$ cross sections. We have demonstrated that the inclusion of the nonperturbative QCD corrections in the photon wave function improves the description of the experimental data for the observables that are sensitive to dipoles of large size, but have negligible impact on the observables that are dominated by short distances. Our results motivate a new global fit of the reduced cross section using the color dipole formalism and taking into account of the npQCD corrections to the photon wave function.

\begin{acknowledgments}
This work was  partially financed by the Brazilian funding
agencies CNPq, CAPES,  FAPERGS and INCT-FNA (process number 
464898/2014-5).
\end{acknowledgments}

\hspace{1.0cm}

\end{document}